\documentclass[12pt]{iopart}

\usepackage{bm}
\usepackage{epsfig}
\usepackage{citesort}
\usepackage{color}
\eqnobysec

\newcommand{\lwig}{\mbox{\;\raisebox{.3ex}
    {$<$}$\!\!\!\!\!$\raisebox{-.9ex}{$\sim$}\;}}

\newcommand{\modif}[1]{\color{blue}}

\begin{document}

\begin{flushright}
{\large \tt MPP-2006-76, \tt LAPTH-1148/06}
\end{flushright}

\title[CMB forecasts from Monte Carlo simulations]%
{Probing cosmological parameters with the CMB:
Forecasts from Monte Carlo simulations}

\author{Laurence Perotto, Julien Lesgourgues}
\address{Laboratoire d'Annecy-le-vieux de Physique Th\'eorique LAPTH
\\ BP110, F-74941 Annecy-le-vieux Cedex, France}

\author{Steen Hannestad, Huitzu Tu}
\address{Department of Physics and Astronomy, University of Aarhus \\
DK-8000 Aarhus C, Denmark}

\author{Yvonne Y. Y. Wong}
\address{Max-Planck-Institut f\"ur Physik
(Werner-Heisenberg-Institut) \\ F\"ohringer Ring 6, D-80805 M\"unchen,
Germany}

\ead{\mailto{perotto@lapp.in2p3.fr},
\mailto{lesgourg@lapp.in2p3.fr}, \mailto{sth@phys.au.dk},
\mailto{huitzu@phys.au.dk}, \mailto{ywong@mppmu.mpg.de}}

\begin{abstract}
The Fisher matrix formalism has in recent times become the standard
method for predicting the precision with which various cosmological
parameters can be extracted from future data.  This approach is fast,
and generally returns accurate estimates for the parameter errors when
the individual parameter likelihoods approximate a Gaussian
distribution.  However, where Gaussianity is not respected (due, for
instance, to strong parameter degeneracies), the Fisher matrix
formalism loses its reliability.  In this paper, we compare the
results of the Fisher matrix approach with those from Monte Carlo
simulations.  The latter method is based on the publicly available
CosmoMC code, but uses synthetic realisations of data sets anticipated
for future experiments.  We focus on prospective cosmic microwave
background (CMB) data from the Planck satellite, with or without CMB
lensing information, and its implications for a minimal cosmological
scenario with eight parameters and an extended model with eleven
parameters. We show that in many cases, the projected sensitivities
from the Fisher matrix and the Monte Carlo methods differ
significantly, particularly in models with many parameters.
Sensitivities to the neutrino mass and the dark matter fraction are
especially susceptible to change.
\end{abstract}

\maketitle

\section{Introduction}

In the last few years precision measurements of the cosmic
microwave background (CMB), large scale structure (LSS) of
galaxies, and distant type Ia supernovae (SNIa) have helped to
establish a new standard model of cosmology. In this model, the
geometry is flat so that $\Omega_{\rm total} = 1$, and the total
energy density is made up of matter ($\Omega_m \sim 0.3$), and
dark energy ($\Omega_\Lambda \sim 0.7$, with equation of state $w
\equiv P/\rho\simeq -1$). With only a few free parameters this
model provides an excellent fit to all current observations.
Furthermore, each of these parameters is very tightly constrained
by the observational data
\cite{Riess:1998cb,Perlmutter:1998np,Astier:2005qq,Spergel:2003cb,Spergel:2006hy,bib:sdss1}.

However, many other parameters can plausibly have a physical
influence on the cosmological data, even if their presence has not
yet been detected. Such parameters are, for example, a running
spectral index of the primordial power spectrum, an equation of
state for the dark energy which differs from $P=-\rho$, and
non-zero neutrino masses (e.g., \cite{Hannestad:2005ey,Lesgourgues:2006nd}).
Indeed, neutrinos are already known to have non-zero masses from
oscillation experiments, where strong evidence points to a mass
in excess of $\sim 0.05$ eV for the heaviest mass eigenstate
(see e.g. \cite{Fogli:2006qg}).
Given the sensitivities of future cosmological probes,
even such a small mass will very likely be measured (e.g., \cite{Hu:1997mj,Hannestad:2002cn,Kaplinghat:2003bh,Lesgourgues:2004ps,Lesgourgues:2005yv,%
Hannestad:2006as,Wang:2005vr,Takada:2005si}).
Thus, for future experiments such as the Planck satellite%
\footnote{ESA home page for the Planck project: {\tt
 http://astro.estec.esa.nl/SA-general/Projects/Planck/} and
 Planck-HFI web site: {\tt http://www.planck.fr/}} and beyond,
it will be
necessary to include the neutrino mass in the data analysis.

When performing a parameter error forecast for future experiments, it is
customary to use the Fisher matrix formalism in which the
formal error bar on a given parameter can be estimated from the
derivatives of the observables with respect to the model parameters
around the best fit point \cite{Tegmark:1996bz,Eisenstein:1998hr}.
However, this approach can only give a reasonable estimation when
the likelihood of the cosmological parameters approximates a
multivariate Gaussian function of the cosmological parameters.  In
practice, significant departures from Gaussianity arise because of
parameter degeneracies (in other words, because specific
combinations of parameters are poorly constrained by the data), or
because a parameter is defined in a finite interval, and
its probability distribution does not fall to zero at one of the boundaries.

In this paper, we explore an alternative, simulation-based
approach to derive projected sensitivities on the cosmological
parameters for future experiments.
This approach utilises synthetic data which emulate the expected
instrumental and observational characteristics of the experiment
under consideration.
Using modern stochastic optimisation methods such
as Markov Chain Monte Carlo or Importance Sampling \cite{Lewis:2002ah},
the projected
parameter errors can be extracted from the synthetic data set
in the same way that parameters are extracted from existing data.
Since this technique makes no assumption about the Gaussianity or
otherwise of the parameter probabilities, we expect it to be
much more reliable than the conventional Fisher matrix forecast.

This simulation-based forecast technique is in principle applicable to any
one cosmological probe or combination of probes of interest,
provided that one can reliably synthesise
the data set given some underlying cosmological model and the
instrumental noise and sensitivity.
In this work, we focus on the example of
the Planck satellite, to be launched in 2007 or 2008 by ESA for measuring
with unprecedented sensitivitiy
the CMB temperature and polarisation anisotropies on the full sky \cite{:2006uk} .

The paper is structured as follows.
In section \ref{sec:CMB_data_model}, we summarise the statistical
properties of the CMB data, with or without lensing information, and
illustrate how synthetic data can be generated for some given
fiducial model and instrumental characteristics.
Sections \ref{sec:Mock} and \ref{sec:Fisher} outline, respectively, the
simulation-based Monte Carlo Markov Chain method and the Fisher matrix
formalism for forecasting cosmological parameter errors.
Section \ref{sec:minimal} contains a detailed comparison of the results
obtained using  these two techniques for a minimal cosmological
model with eight independent parameters.  The analysis is extended
in section \ref{sec:non-minimal} to a more complicated model with
eleven  parameters, a case in which the differences between
the Monte Carlo and Fisher matrix results are exacerbated.  We provide our
conclusions in section\ \ref{sec:Conclusions}.

\section{CMB data model} \label{sec:CMB_data_model}

Raw data from a CMB probe such as the Planck satellite
can be optimally reduced to sky maps
for the three observables of interest: the temperature and the
two polarisation modes~\cite{Kamionkowski:1996ks}.  Maps are usually expanded in spherical
harmonics, where the coefficients, or multipole moments, $a_{lm}$ receive contributions
from both the CMB signal $s_{lm}$ and the experimental noise $n_{lm}$,
\begin{equation}
a_{lm}^P = s_{lm}^P +  n_{lm}^P.
\end{equation}
Here, the index $P$ runs over $T$ (temperature), $E$ (curl-free polarisation),
and $B$ (divergence-free polarisation).
The noise part can be modelled as a combined effect of a Gaussian beam
and a spatially uniform Gaussian white noise.  Thus, for an experiment with some known
beam width and sensitivity, the noise power spectrum can be approximated as
\begin{equation}
\label{eq:tebnoise}
N_l^{PP'} \equiv \langle n_{lm}^{P*} n_{lm}^{P'} \rangle =
\delta_{P P'} \
\theta_\mathrm{fwhm}^2 \
\sigma_P^2 \ \exp\left[l(l+1) \ \frac{\theta_\mathrm{fwhm}^2}{8 \ln 2}\right],
\end{equation}
where $\theta_\mathrm{fwhm}$ is the full width at half maximum of the
Gaussian beam, and $\sigma_P$ is the root mean square of the
instrumental noise.  Non-diagonal noise terms (i.e., $P \neq P'$)
 are expected to vanish since the noise contributions
from different maps are uncorrelated.  The assumption of a spatially
uniform Gaussian noise spectrum ensures that the noise term is diagonal
in the $l$ basis.

The signal $s_{lm}^P$ contains {\it ab initio} various contributions
from the primary anisotropies (related to primordial inhomogeneities on
the last scattering surface), the secondary anisotropies (caused by the
interaction of primary CMB photons with the intervening medium), and a
wide variety of astrophysical foregrounds~\cite{Hu:1995em}.  Since foreground emissions
typically have non-thermal spectra (except for a small contribution of the
kinetic Sunyaev--Zel'dovitch effect), they can be accurately
removed by combining data from various frequency bands.  After
foreground cleaning, CMB maps should contain only the primary anisotropies,
plus a few secondary effects
such as the late integrated Sachs--Wolfe (ISW) effect and weak lensing
distortions~\cite{Ber97,LE,LE2,Lewis:2006fu}
(both caused by the neighbouring distribution of dark matter
and baryons) which do not alter the CMB's Planckian shape.
For the temperature and the $E$-polarisation mode, the effect of
weak lensing on $s_{lm}^P$ is small and can be neglected in a first approximation
(this is not true for the $B$-mode, at least on small angular scales,
or, equivalently, at large multipoles).  Thus, for a
full-sky experiment, each multipole moment $s_{lm}^T$ and $s_{lm}^E$ is an
independent Gaussian variable. Since the signal is also uncorrelated with
the noise, the power spectrum of the total $a_{lm}$
(after foreground cleaning) reads
\begin{equation}
\langle a_{lm}^{P*}a_{l'm'}^{P'}\rangle =
\left(C_l^{PP'}+N_l^{PP'}\right) \delta_{ll'}\delta_{mm'}~,
\end{equation}
where the Dirac delta functions ensure that different $l$ and $m$ modes are
uncorrelated.

A number of methods are available on the market for the extraction of the weak
lensing deflection angle ${\bf d} ( {\bf n} )$ from the CMB signals;
the role of weak lensing is to remap the
direction of observation from ${\bf n}$ to ${\bf n}'={\bf n}+{\bf d} (
{\bf n} )$~\cite{IE,Lewis:2006fu}.  These extraction methods exploit the
non-Gaussian properties of the signal $s_{lm}^P$ induced by lensing.
The reconstructed deflection field can be specified by a single set of
expansion coefficients $a_{lm}^d$ in harmonic space,
since in a first approximation the vector field ${\bf d} ( {\bf n} )$ is
curl-free~\cite{IE}.  The ${\bf d} ( {\bf n} )$ field itself becomes
non-Gaussian at low redshifts due to the non-linear evolution of the
gravitational potential.  However, at sufficiently large angular scales (i.e.,
$l \lwig 1000$), contributions to the deflection field will come mainly from
the linear regime.  Thus, $a_{lm}^d$
can be considered as an approximately Gaussian variable~\cite{Okamoto2003}.

The power spectrum of the deflection field reads
\begin{equation}
\langle a_{lm}^{d*} a_{l'm'}^{d}\rangle =
\left(C_l^{dd}+N_l^{dd}\right) \delta_{ll'}\delta_{mm'},
\end{equation}
where the noise power spectrum $N_l^{dd}$  reflects the errors in the
deflection map reconstruction, and can be estimated for a given
combination of lensing extraction technique and experiment.
Here, we refer to the
{\it quadratic estimator method} of Hu \& Okamoto~\cite{Okamoto2003}, 
which provides five estimators of $a_{lm}^d$ based
on the correlations between five possible pairs of maps:
$TT$, $EE$, $TE$, $TB$, $EB$. The estimator $BB$ (from
self-correlations in the $B$-mode map) cannot be used in this method,
because the $B$-mode signal is dominated by lensing on small scales.
The authors of \cite{Okamoto2003} also provide
an algorithm for estimating $N_l^{dd}$ given some hypothetical
observed power spectra $C_l^{PP'} + N_l^{PP'}$.  
This final noise power spectrum $N_l^{dd}$ corresponds to the minimal
noise spectrum achievable by optimally combining the five quadratic estimators.
Note that the $B$-mode can potentially play a crucial role here, since it is
the only observable that presents a clear lensing signal.  This is why,
for a sufficiently sensitive experiment, the $EB$-correlation will provide
the best quadratic estimator.  At the precision level of Planck, however,
most of the sensitivity to lensing will come from the $TT$ estimator~\cite{Hu:2001kj,Lesgourgues:2005yv}.

Finally, there exists some non-vanishing correlations between the
temperature and the deflection maps,
\begin{equation}
\langle a_{lm}^{T*} a_{l'm'}^{d}\rangle = C_l^{Td}
\delta_{ll'}\delta_{mm'}~.
\end{equation}
Indeed, the temperature map includes
the well-known ISW effect, a secondary anisotropy induced by the
time-evolution of the gravitational potential wells
during dark energy domination.  The same potential wells are also
responsible for the weak lensing distortions.

Given a fiducial cosmological model, one can use a Boltzmann code
such as {\sc CAMB}\footnote{\tt
http://camb.info/}~\cite{Lewis:1999bs} to calculate the power
spectra $C_l^{TT}$, $C_l^{TE}$, $C_l^{EE}$, $C_l^{BB}$, $C_l^{dd}$,
$C_l^{Td}$. Together with the noise spectra
$N_l^{TT}$, $N_l^{EE}=N_l^{BB}$, and $N_l^{dd}$ from equation (\ref{eq:tebnoise}) and
the algorithm of \cite{Okamoto2003}, one can generate synthetic data with the appropriate correlations and
noise characteristics using the following procedure:

\begin{enumerate}
\item Generate Gaussian-distributed random numbers $G_{lm}^{(i)}$ with unit variance.

\item Define the $T$, $E$ and $d$ multipole moments as
\begin{eqnarray}
a_{lm}^T &=& \sqrt{\bar{C}_l^{TT}} G_{lm}^{(1)},\nonumber\\
a_{lm}^E &=& \frac{\bar{C}_{l}^{TE}}{\bar{C}_{l}^{TT}}\sqrt{\bar{C}^{TT}_{l}}
G^{(1)}_{lm} +
\sqrt{\bar{C}_{l}^{EE}-\frac{(\bar{C}_{l}^{TE})^2}{\bar{C}_{l}^{TT}}}
G^{(2)}_{lm}, \label{realisation}\\
a_{lm}^d &=& \frac{\bar{C}_{l}^{Td}}{\bar{C}_{l}^{TT}}\sqrt{\bar{C}^{TT}_{l}}
G^{(1)}_{lm} +
\sqrt{\bar{C}_{l}^{dd}-\frac{(\bar{C}_{l}^{Td})^2}{\bar{C}_{l}^{TT}}}
G^{(3)}_{lm}, \nonumber
\end{eqnarray}
where $\bar{C}_l^{PP'} \equiv C_l^{PP'}+N_l^{PP'}$ is the fiducial
signal plus noise spectrum.

\item Given a realisation of the $a_{lm}^P$'s, we can estimate
the power spectra of the mock data by
\begin{equation}
\label{reconstruction}
\hat{C}_l^{PP'} = \frac{1}{2l+1}\left( a_{l0}^{P}a_{l0}^{P'}
+ 2\sum_{m=0}^{l}a_{lm}^{P*}a_{lm}^{P'}\right)~.
\end{equation}

\end{enumerate}
Note that we do not discuss the simulation of the $B$-mode
polarisation, because they are not relevant for the analysis presented
in this work.  Indeed, the measurement of the $B$-mode by Planck is
expected to be noise-dominated rather than cosmic-variance-dominated.
Thus, unless one wants to constrain the amplitude of primordial
gravitational waves, the $B$-mode can be safely neglected for
parameter extraction from Planck.

\section{Monte Carlo analysis of the mock data} \label{sec:Mock}

In order to extract the cosmological parameter errors from
the mock Planck data, we perform a Bayesian likelihood analysis.

In our model, each data point has contributions from both signal and
noise. Since both contributions are Gaussian-distributed, one can write the
likelihood function of the data given the theoretical model
as~\cite{Tegmark:1996bz}
\begin{equation}
{\cal L}({\mathbf a} |\Theta) \propto \exp \left( -\frac{1}{2} \mathbf{a}^\dagger
[\bar{C}(\Theta)^{-1}] \mathbf{a} \right),
\label{likelihood}
\end{equation}
where $\mathbf{a} = \{a^T_{lm},a^E_{lm},a^d_{lm} \}$ is the data
vector,
$\Theta = (\theta_1, \theta_2, \ldots)$ is a vector describing
the theoretical model parameters, and $\bar{C}(\Theta)$ is the
theoretical data covariance matrix (cf.\ the mock data
covariance matrix, $\hat{C} \equiv \langle \mathbf{a} 
\mathbf{a}^\dagger \rangle$).  The
maximum likelihood is an unbiased estimator, i.e.,
\begin{equation}
\langle \Theta \rangle = \Theta^0,
\end{equation}
where $\Theta^0$ indicates the parameter vector of the underlying
cosmological model, $\Theta$ is the one reconstructed by
maximising the likelihood (i.e., the so-called best-fit model),
and $\langle \ldots \rangle$ denotes an average over many
independent realisations.

The probability
distribution for each parameter or a subset of parameters can be
reconstructed using Bayes theorem.  If we assume flat priors on the
parameters $\theta_i$, the distribution is simply obtained by
integrating the likelihood along unwanted parameters, a process
called marginalisation. Confidence levels for each parameter are
then defined as the regions in which the probability exceeds a given
value. If we are interested only in these confidence level, it is
straightforward to show that the normalisation factor in front of
the likelihood function (\ref{likelihood}) is irrelevant.
In other words, the effective $\chi^2$,
$\chi_{\rm eff}^2 \equiv - 2 \ln {\cal L}$, can be shifted by
an arbitrary constant without changing the results. The effective
$\chi^2$ can be derived from (\ref{likelihood}),
\begin{equation}
\chi^2_{\rm eff} = \sum_{l} (2l+1) \left( \frac{D}{|\bar{C}|} +
\ln{\frac{|\bar{C}|}{|\hat{C}|}} - 3 \right),
\end{equation}
where $D$ is defined as
\begin{eqnarray}
D  &=&
\hat{C}_l^{TT}\bar{C}_l^{EE}\bar{C}_l^{dd} +
\bar{C}_l^{TT}\hat{C}_l^{EE}\bar{C}_l^{dd} +
\bar{C}_l^{TT}\bar{C}_l^{EE}\hat{C}_l^{dd} \nonumber\\
&&- \bar{C}_l^{TE}\left(\bar{C}_l^{TE}\hat{C}_l^{dd} +
2\hat{C}_l^{TE}\bar{C}_l^{dd} \right)
- \bar{C}_l^{Td}\left(\bar{C}_l^{Td}\hat{C}_l^{EE} +
2\hat{C}_l^{Td}\bar{C}_l^{EE} \right),
\end{eqnarray}
and $|\bar{C}|$, $|\hat{C}|$ denote the determinants of
the theoretical and observed data covariance matrices respectively,
\begin{eqnarray}
|\bar{C}| &=& \bar{C}_l^{TT}\bar{C}_l^{EE}\bar{C}_l^{dd} -
\left(\bar{C}_l^{TE}\right)^2\bar{C}_l^{dd} -
\left(\bar{C}_l^{Td}\right)^2\bar{C}_l^{EE} ~, \\
|\hat{C}| &=& \hat{C}_l^{TT}\hat{C}_l^{EE}\hat{C}_l^{dd} -
\left(\hat{C}_l^{TE}\right)^2\hat{C}_l^{dd} -
\left(\hat{C}_l^{Td}\right)^2\hat{C}_l^{EE}~.
\end{eqnarray}
 In these expressions, the arbitrary normalisation term has been
chosen in such way that $\chi^2_{\rm eff}=0$ if
$\bar{C}_l^{PP'}=\hat{C}_l^{PP'}$.

 All expressions introduced so
far assume full sky coverage; real experiments, however, can only
see a fraction of the sky. Even for satellite experiments a map cut must be
performed in order to eliminate point sources and 
galactic plane foreground
contaminations. As a result, different multipole moments $a_{lm}^P$
(but in the same mode $P$) become correlated with each other.
The likelihood function
in this case takes a rather complicated form, depending on the shape
of the remaining observed portion of the sky. However, for experiments
probing almost the full sky (e.g., COBE, WMAP, or Planck), correlations
are expected only between neighbouring multipoles. In order to simplify
the problem, one can take the $a_{lm}^P$'s to be uncorrelated, and
introduce a factor
$f_{\rm sky}$ in the effective $\chi^2$,
\begin{equation}
\chi^2_{\rm eff} = \sum_{l} (2l+1) f_{\rm sky} \left(
\frac{D}{|\bar{C}|} + \ln{\frac{|\bar{C}|}{|\hat{C}|}} - 3 \right),
\label{chieff}
\end{equation}
where $f_{\rm sky}$ denotes the observed fraction of the sky.  In other
words, instead of measuring $(2l+1)$ independent moments at each value of $l$,
the number of degrees of freedom is now reduced to $(2l+1)f_{\rm sky}$.
It is possible to build better approximations~\cite{Wandelt:2000av},
but for simplicity we will model the Planck data in this way
with $f_{\rm sky}=0.65$,
corresponding roughly to what remains after a sky cut has been imposed
near the galactic plane.

Given some mock data set, it is straightforward to sample the likelihood
and estimate the marginalised probability distribution using, for example,
the publicly available code CosmoMC\footnote{\tt
http://cosmologist.info/cosmomc/}~\cite{Lewis:2002ah}, which explores the parameter space
$\Theta$ by means of Monte Carlo Markov Chains. The interface between data
and model should make use of equation (\ref{chieff}), and should
require minimal modifications to the public version of CosmoMC.
Indeed, CosmoMC already contains a subroutine called {\tt ChiSqExact},
which uses an expression for $\chi^2_{\rm eff}$
equivalent to our (\ref{chieff})
in the absence of lensing information.

\section{The Fisher matrix analysis} \label{sec:Fisher}

The Fisher matrix technique allows for a quick, analytic estimate of
the confidence limits by approximating the likelihood function
${\mathcal L}({\mathbf a}|\Theta)$ as a multivariate Gaussian
function of the theoretical parameters $\Theta$.  Since
${\mathcal L}({\mathbf a}|\Theta)$ is generally a rather complicated
function of $\Theta$, this approximation will likely lead to incorrect
results.  Indeed, the goal of this paper is to determine in
concrete cases the precision of the Fisher matrix analysis
compared with the Monte Carlo approach described in the last section.

The likelihood function should peak at $\Theta \simeq \Theta^0$,
and can be Taylor expanded to second order around this value. The
relevant term at second order is the Fisher information matrix,
defined as
\begin{equation}
F_{ij}\equiv
\left.
- \frac{\partial^2 \ln {\mathcal L}}{\partial \theta_i \partial \theta_j}
\right|_{\, \Theta^0}~.
\label{fisher_def}
\end{equation}
The Fisher matrix is closely related to the precision with which
the parameters $\theta_i$ can be constrained. If all free
parameters are to be determined from the data alone with flat
priors, then it follows from the Cramer--Rao inequality
\cite{kendall} that the formal error on the parameter $\theta_i$ is given by
\begin{equation}
\label{eq:fishererror}
\sigma(\theta_i) = \sqrt{(F^{-1})_{ii}}
\end{equation}
for an optimal unbiased estimator such as the maximum likelihood
\cite{Tegmark:1996bz}.

Plugging equations (\ref{likelihood}) and (\ref{chieff}) into the above expression,
one finds
\begin{equation}
F_{ij} = \sum_{l=2}^{l_{\rm max}}\sum_{PP',QQ'}\frac{\partial
C_l^{PP'}}{\partial
  \theta_i}({\rm Cov}_l^{-1})_{PP'QQ'}\frac{\partial C_l^{QQ'}}{\partial \theta_j},
\label{cov}
\end{equation}
where $l_{\rm max}$ is the maximum multipole available given the
angular resolution of the considered experiment, and
$PP',QQ' \in \{TT,EE,TE,dd,Td\}$.
The matrix ${\rm Cov}_l$ is the power spectrum covariance
matrix at the $l$-multipole,
\begin{eqnarray}
{\rm Cov}_l=\frac{2}{(2l+1)f_{\rm sky}}\left(\begin{array}{ccccc}
\Xi_{TTTT} & \Xi_{TTEE} & \Xi_{TTTE} & \Xi_{TTTd} & \Xi_{TTdd} \\
\Xi_{TTEE} & \Xi_{EEEE} & \Xi_{TEEE} & 0 & 0\\
 \Xi_{TTTE} & \Xi_{TEEE} & \Xi_{TETE} & 0 & 0\\
 \Xi_{TTTd}& 0 & 0 & \Xi_{TdTd} & \Xi_{Tddd}\\
\Xi_{TTdd} & 0 & 0 & \Xi_{Tddd} & \Xi_{dddd} \\
\end{array} \right),
\end{eqnarray}
where the auto-correlation coefficients are given by
\begin{eqnarray}
\Xi_{TTTT}&=&
\left[\left(\bar{C}_l^{TT}\right)^2-\frac{2\left(\bar{C}_l^{TE}\right)^2
\left(\bar{C}_l^{Td}\right)^2}{\bar{C}_l^{EE}\bar{C}_l^{dd}}\right],\nonumber\\
\Xi_{EEEE}&=& \left(\bar{C}_l^{EE}\right)^2,\nonumber\\
\Xi_{TETE}&=& \frac{1}{2}\left[ \left(\bar{C}_l^{TE}\right)^2+
\bar{C}_l^{TT}\bar{C}_l^{EE}\right] -
\frac{\bar{C}_l^{EE}\left(\bar{C}_l^{Td}\right)^2}{2\bar{C}_l^{dd}},\nonumber\\
\Xi_{TdTd}&=& \frac{1}{2}\left[ \left(\bar{C}_l^{Td}\right)^2 +
\bar{C}_l^{TT}\bar{C}_l^{dd}\right] -
\frac{\bar{C}_l^{dd}\left(\bar{C}_l^{TE}\right)^2}{2\bar{C}_l^{EE}},\nonumber\\
\Xi_{dddd}&=& \left(\bar{C}_l^{dd}\right)^2,
\end{eqnarray}
while the cross-correlated ones are
\begin{eqnarray}
\Xi_{TTEE} &=& \left(\bar{C}_l^{TE}\right)^2,\nonumber\\
\Xi_{TTTE} &=& \bar{C}_l^{TE}\left[ \bar{C}_l^{TT} -
\frac{\left(\bar{C}_l^{Td}\right)^2}{\bar{C}^{dd}_l}\right],\nonumber\\
\Xi_{TEEE} &=& \bar{C}_l^{TE}\bar{C}_l^{EE},\nonumber\\
\Xi_{TTdd} &=& \left(\bar{C}_l^{Td}\right)^2,\nonumber\\
\Xi_{TTTd} &=& \bar{C}_l^{Td}\left[ \bar{C}_l^{TT} -
\frac{\left(\bar{C}_l^{TE}\right)^2}{\bar{C}_l^{EE}}\right],\nonumber\\
\Xi_{Tddd} &=& \bar{C}_l^{Td}\bar{C}_l^{dd}.
\end{eqnarray}

\begin{figure*}[t]
\begin{center}
\includegraphics[width=.31\textwidth]{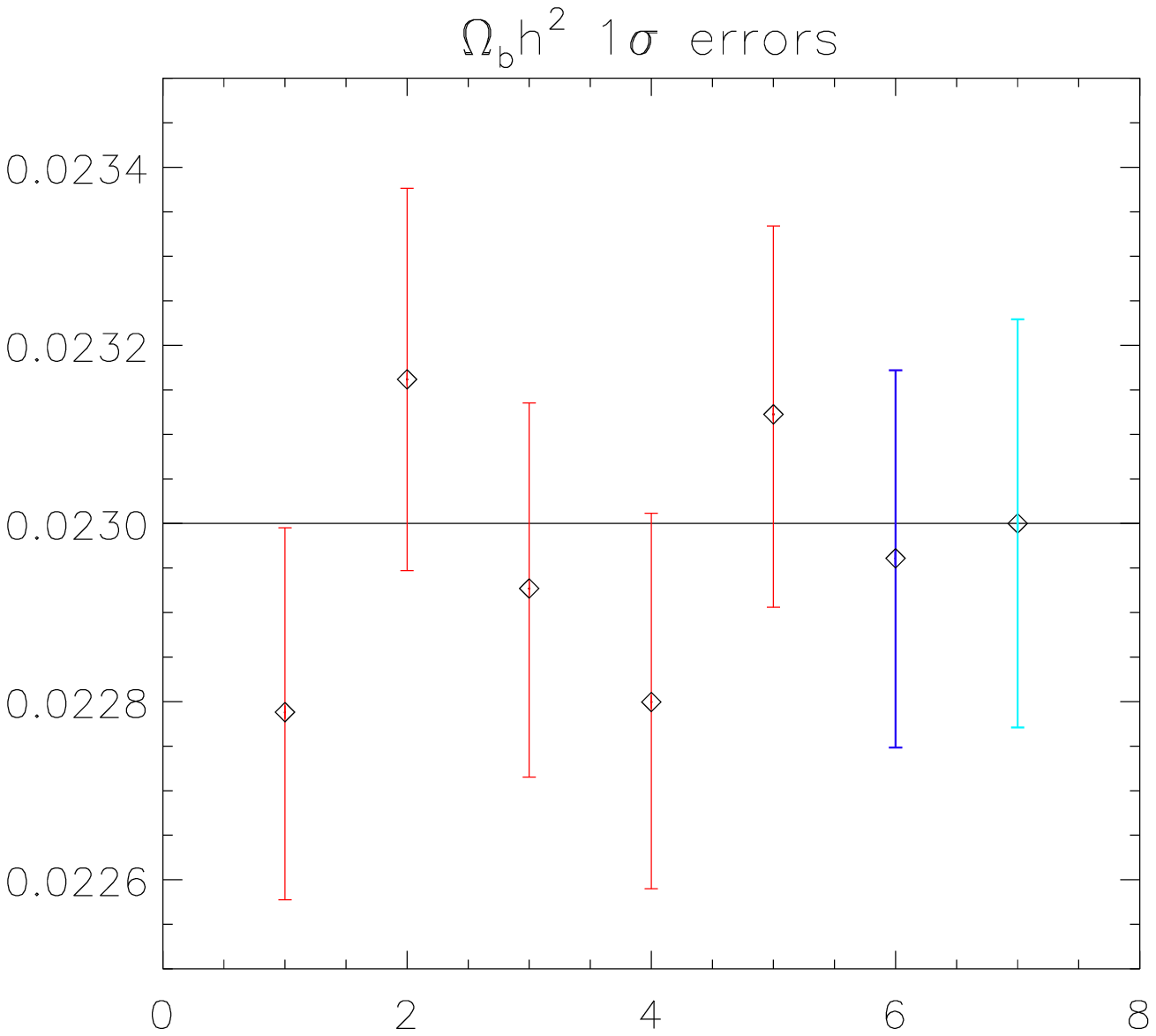}
\hfill
\includegraphics[width=.31\textwidth]{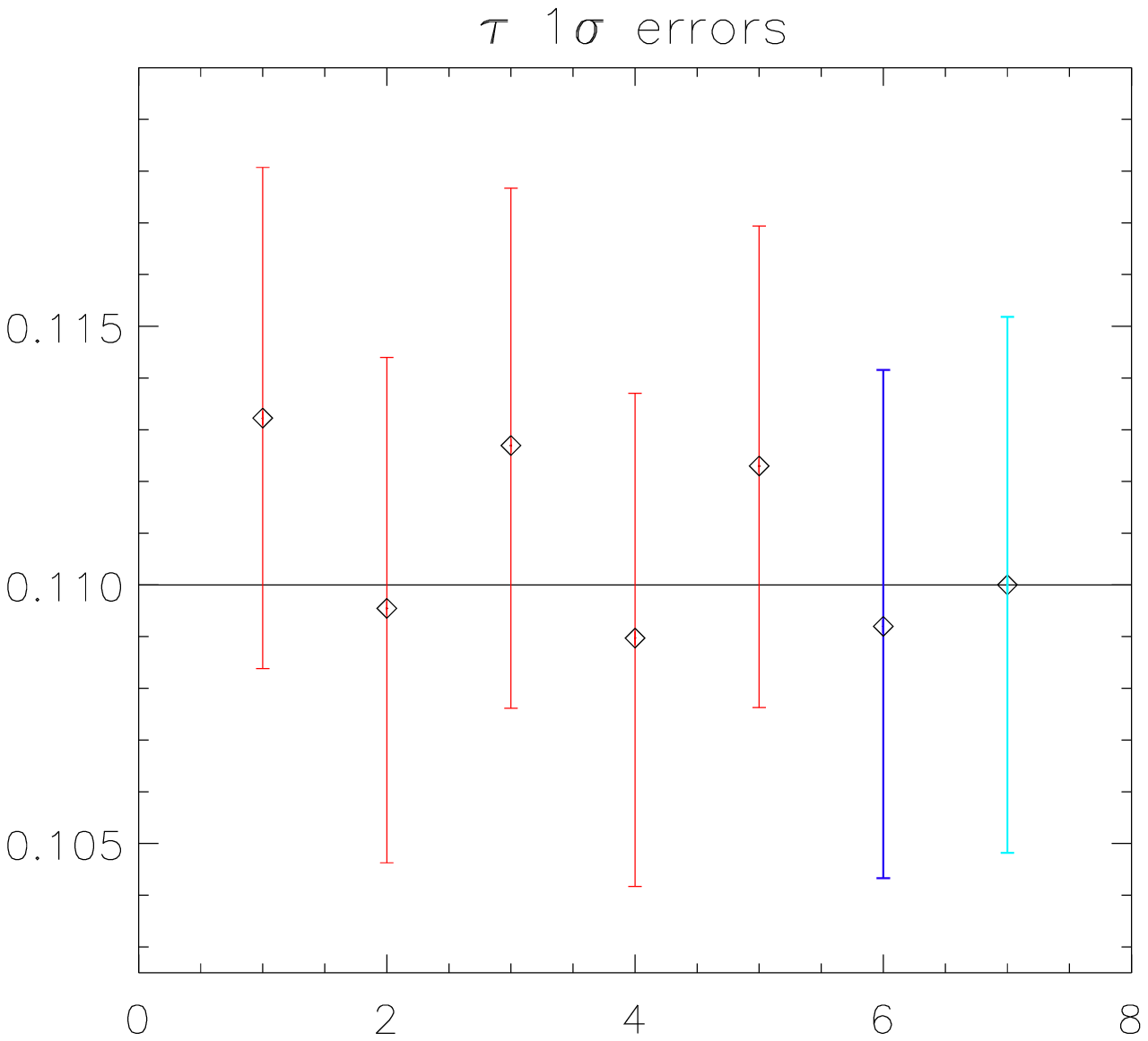}
\hfill
\includegraphics[width=.31\textwidth]{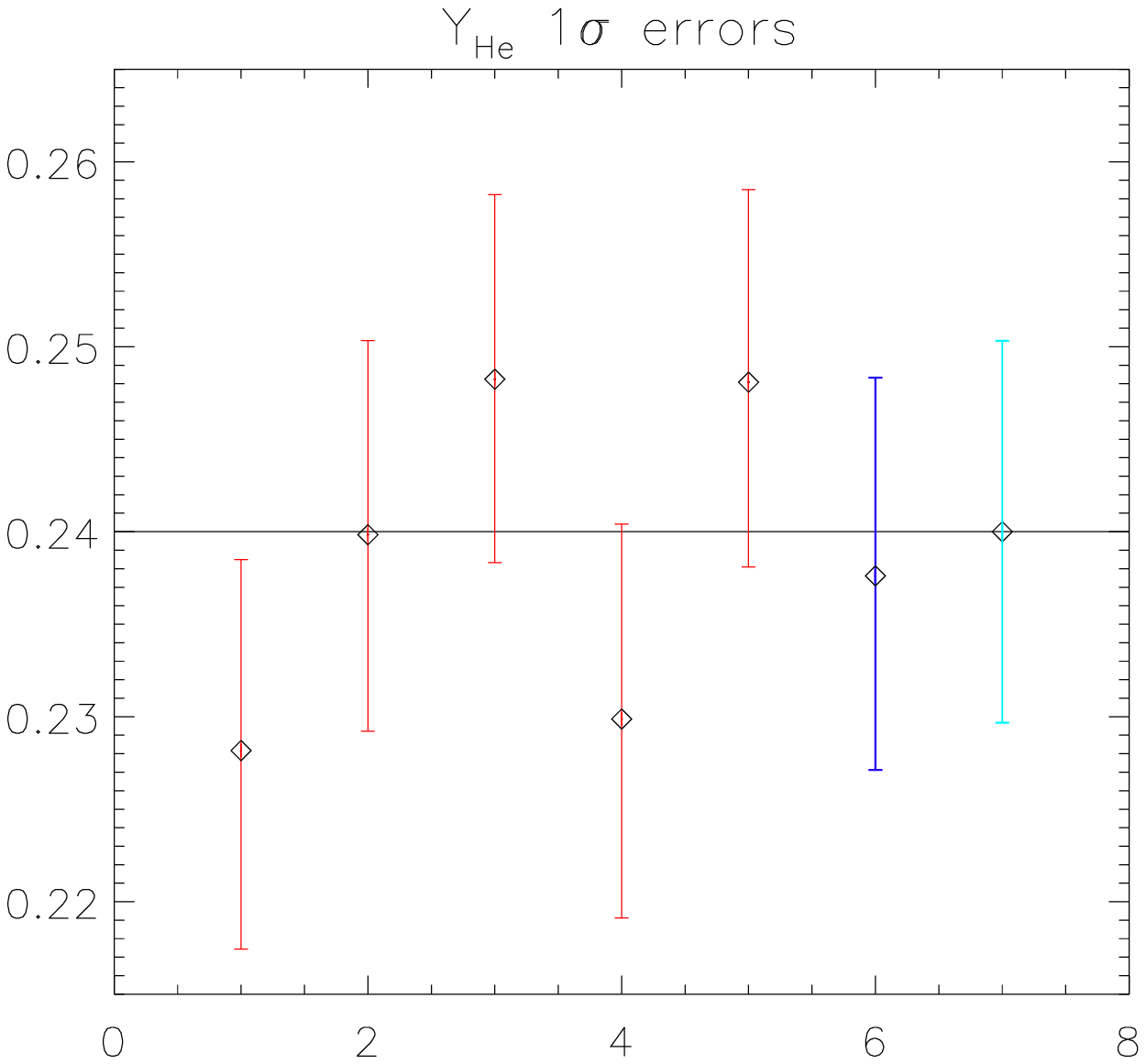}
\hfill
\includegraphics[width=.31\textwidth]{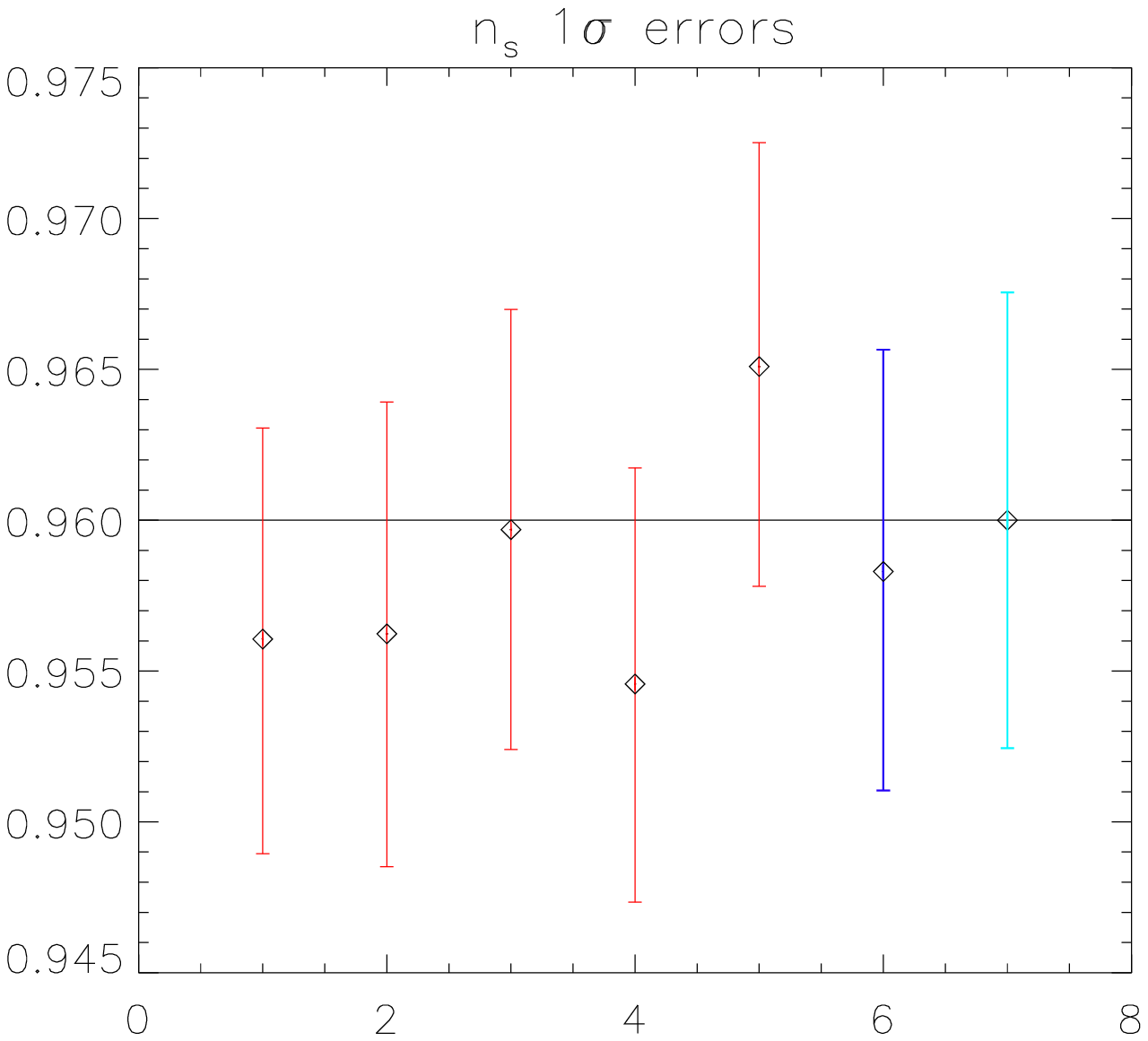}
\hfill
\includegraphics[width=.31\textwidth]{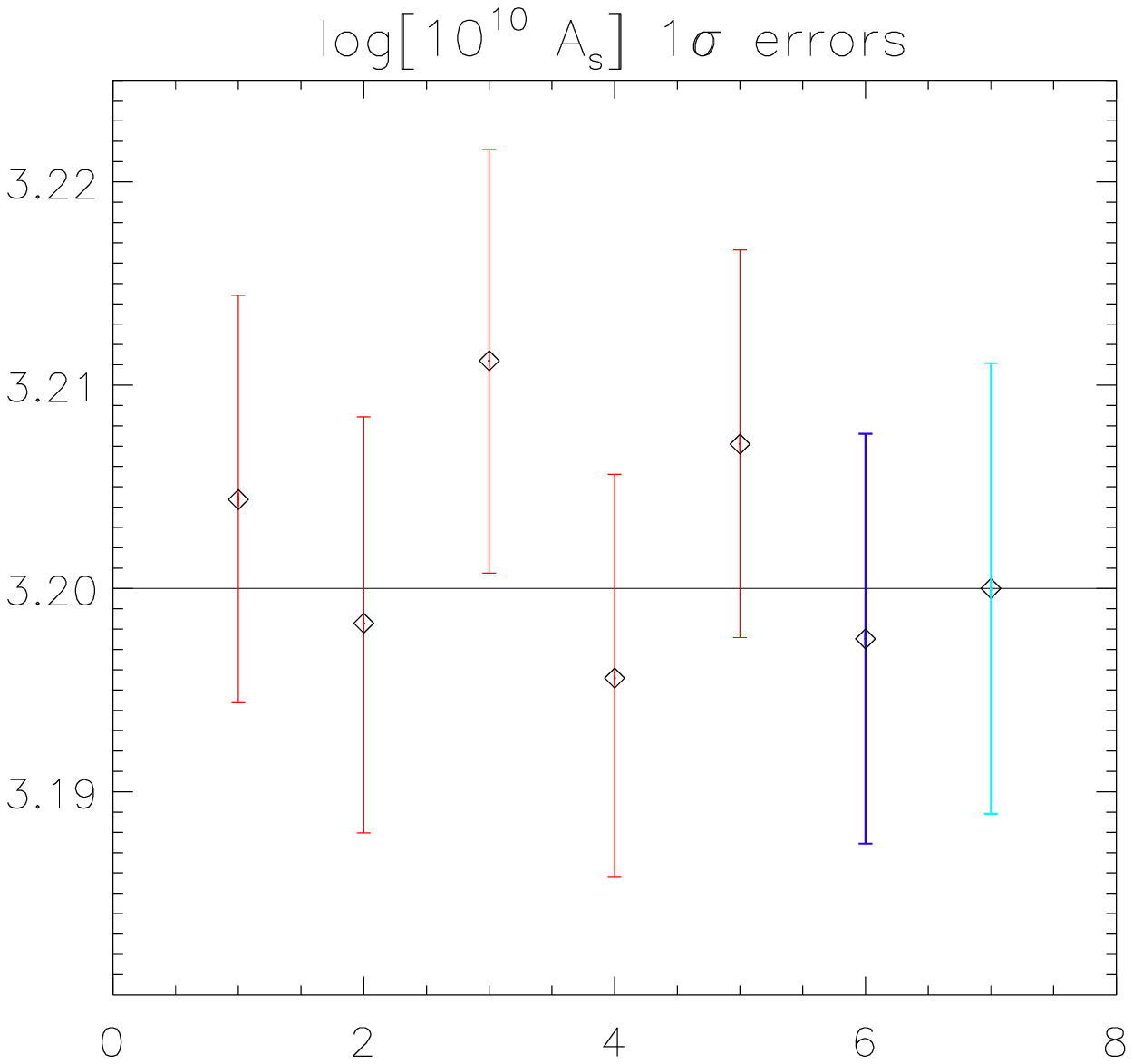}
\hfill
\includegraphics[width=.31\textwidth]{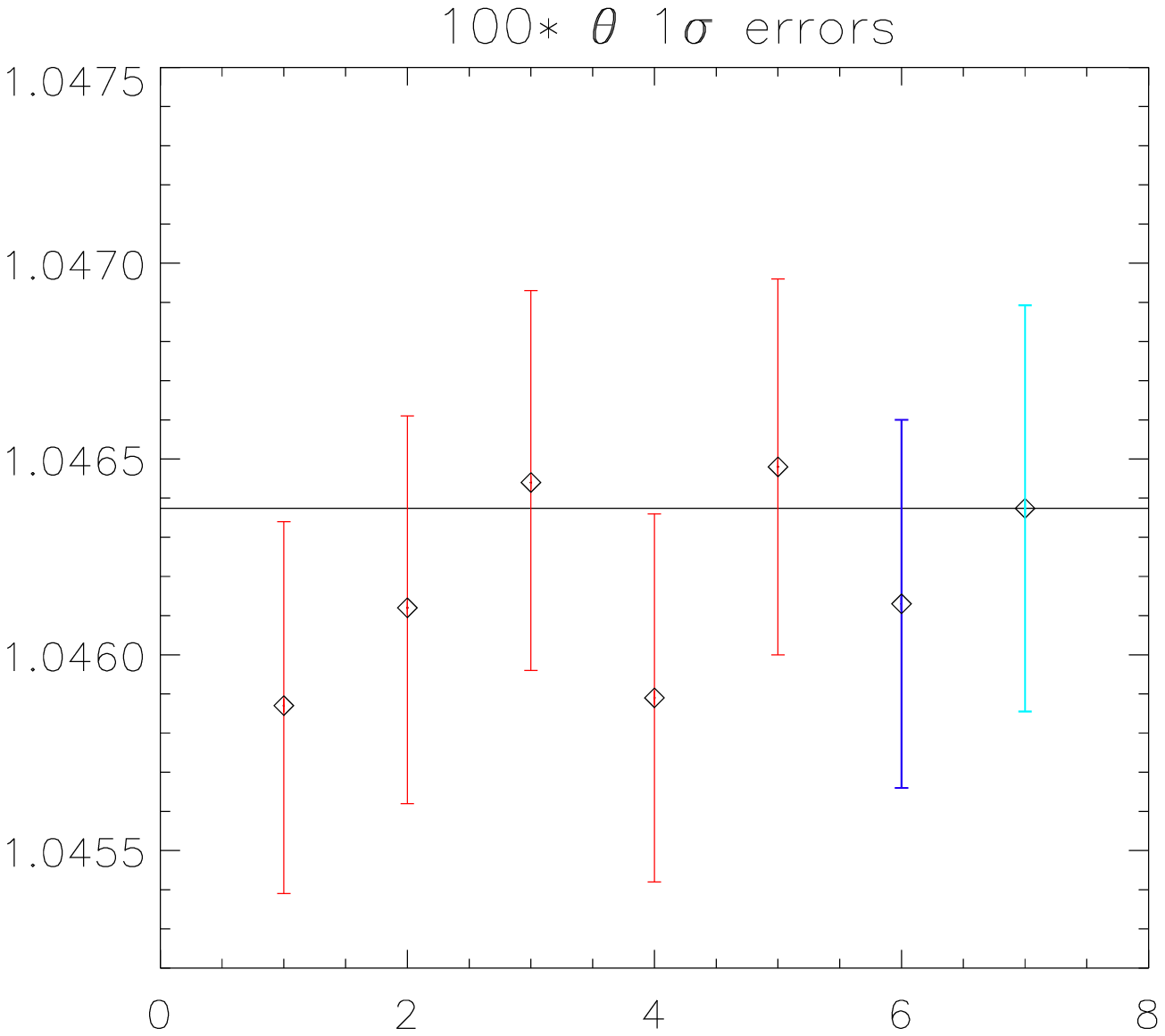}
\hfill
\caption{\label{fig1} Projected 1$\sigma$ errors for
$\{\Omega_bh^2, \tau, Y_{He}, n_s, A_s, \theta \}$ in
the eight-parameter model of section \ref{sec:minimal}.
The first five points and error bars (red) in each plot are the Monte
Carlo estimates from independent mock datasets. The sixth one (dark blue) is
obtained by replacing the mock data spectra by the fiducial spectra.
The last error bar (light blue) corresponds to the
estimate from the Fisher matrix method, centred on the
fiducial value of the parameter of interest (horizontal lines).}
\end{center}
\end{figure*}

The advantage of the Fisher matrix technique is that it is
computationally tractable, and involves much
less numerical machinery than a Markov Chain Monte Carlo
exploration of the parameter space.
However, we emphasise that this method is not equivalent to a full
likelihood analysis. This is because the Taylor expansion is valid
only in regions close to the best fit point.
As we move away from this point the errors can become
larger or smaller than the error (\ref{eq:fishererror}),
depending on the sign of, e.g., the skewness and
kurtosis of the full likelihood function.
Furthermore, the Fisher matrix is sensitive to small numerical
errors in the computation of the derivatives $\partial C^{PP'}_l/\partial
\theta_i$, and elements that are close to zero can be
amplified significantly when inverting the matrix. This often
leads to artificial reduction in the estimated errors, a point
discussed in detail in, for instance, reference \cite{Eisenstein:1998hr}.

Concretely, these issues are related to the strategy for
evaluating numerically the derivatives
${\partial C_l^{PP'}}/{\partial \theta_i}$. Whenever possible, one
should compute a two-sided finite difference $[C_l^{PP'}
(\theta_i^0+\Delta \theta_i)- C_l^{PP'} (\theta_i^0-\Delta
\theta_i)]/(2 \Delta \theta_i)$. The usual prescription for
computing derivatives is to choose as small a stepsize
$\Delta \theta_i$ as possible without introducing too much numerical
noise in the result.
In practice, one can increase $\Delta
\theta_i$ until the derivatives are smooth and exempt of violent
oscillations introduced by numerical instability.

However, this prescription is not necessarily the best choice in
the present context. In order to output, for instance, a reliable
estimate of the 68\% confidence limit (C.L.) $\sigma(\theta_i)$, a
better approach might be to use a stepsize of order $\Delta \theta_i \sim
\sigma(\theta_i)$.  This choice of stepsize should, hopefully,
provide an appropriate average over the 68\% confidence region.
However, there is no well-controlled method to check the
consistency of this approach, short of performing the full
likelihood calculation. It also does not properly allow for
treating parameter correlations.

\section{Comparision of the two methods for a minimal model}
\label{sec:minimal}

We now compare the Fisher matrix and Monte Carlo methods for the case of the
simplest possible cosmological model (in the sense that each parameter
describes a physical effect which is known to occur, and to which
Planck is potentially sensitive). This model is a flat, adiabatic
$\Lambda$ mixed dark matter ($\Lambda$MDM) model with no tensor contributions,
parameterised by the quantities
$\{\Omega_bh^2, \Omega_{dm}h^2, f_{\nu}, \Omega_{\Lambda}, \tau,
Y_{He}, A_s, n_s \}$, representing respectively the baryon and the dark
matter densities, the hot fraction of dark matter $f_{\nu}\equiv
\Omega_{\nu}/\Omega_{dm}$, the cosmological constant energy density,
the optical depth to reionisation, the primordial
Helium fraction, and, finally, the primordial spectrum amplitude and
spectral index.

In order to perform a neat parameter extraction, one should
choose a parameter basis which minimises the parameter correlations.
Thus the choice of the basis should be dictated by an
analysis of all the physical effects on the CMB observables.
The authors of \cite{Eisenstein:1998hr} stress that it is particularly useful to employ, in place of
$\Omega_{\Lambda}$ (or $H_0$), the angular
diameter $\theta$ of the sound horizon at decoupling,
since $\theta$ can be very well determined by the position of the
first acoustic peak.  On the other hand, $\Omega_{\Lambda}$
(or $H_0$) is usually correlated with other variables.

\begin{figure*}[t]
\begin{center}
\includegraphics[width=.45\textwidth]{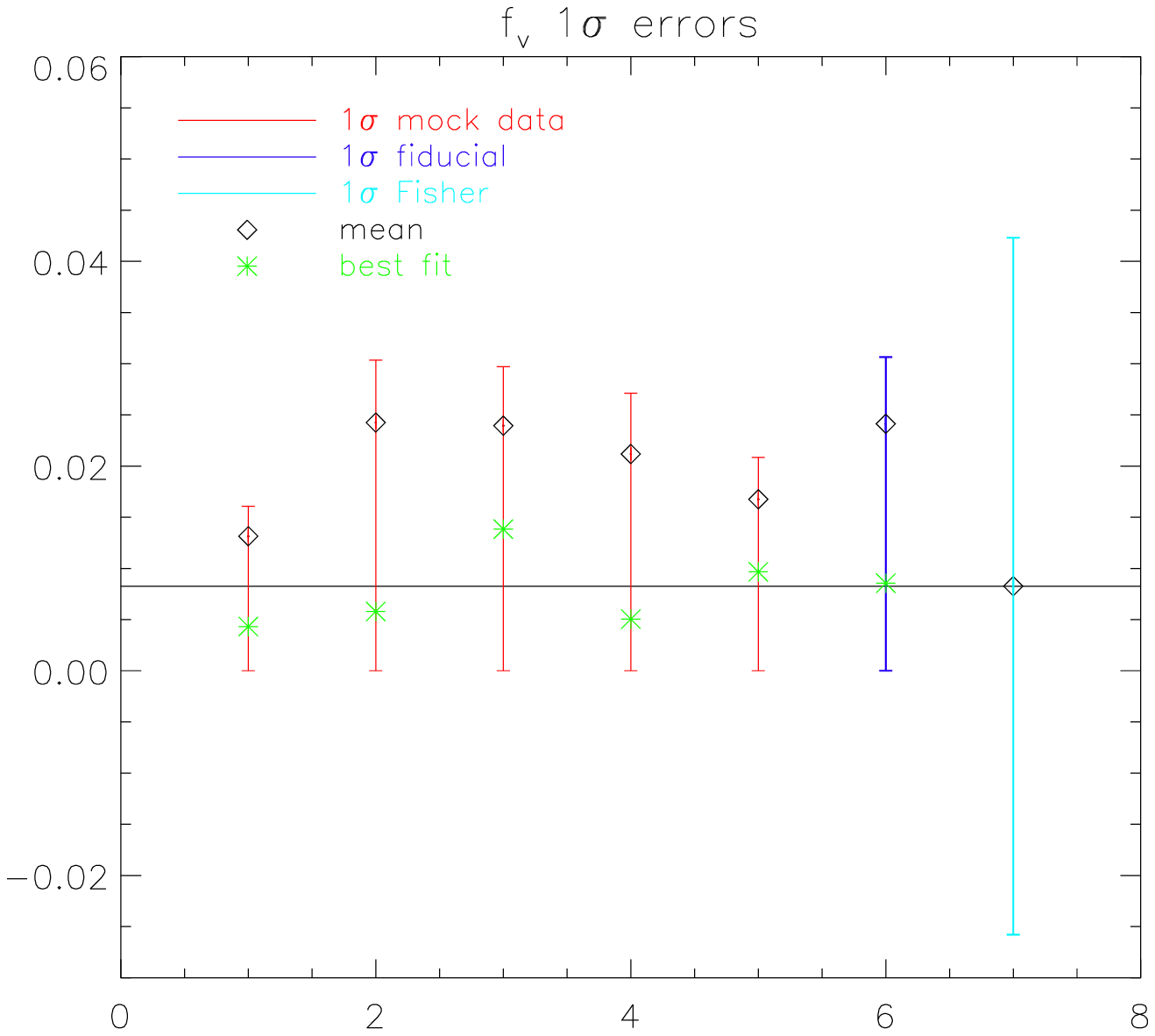}
\hfill
\includegraphics[width=.45\textwidth]{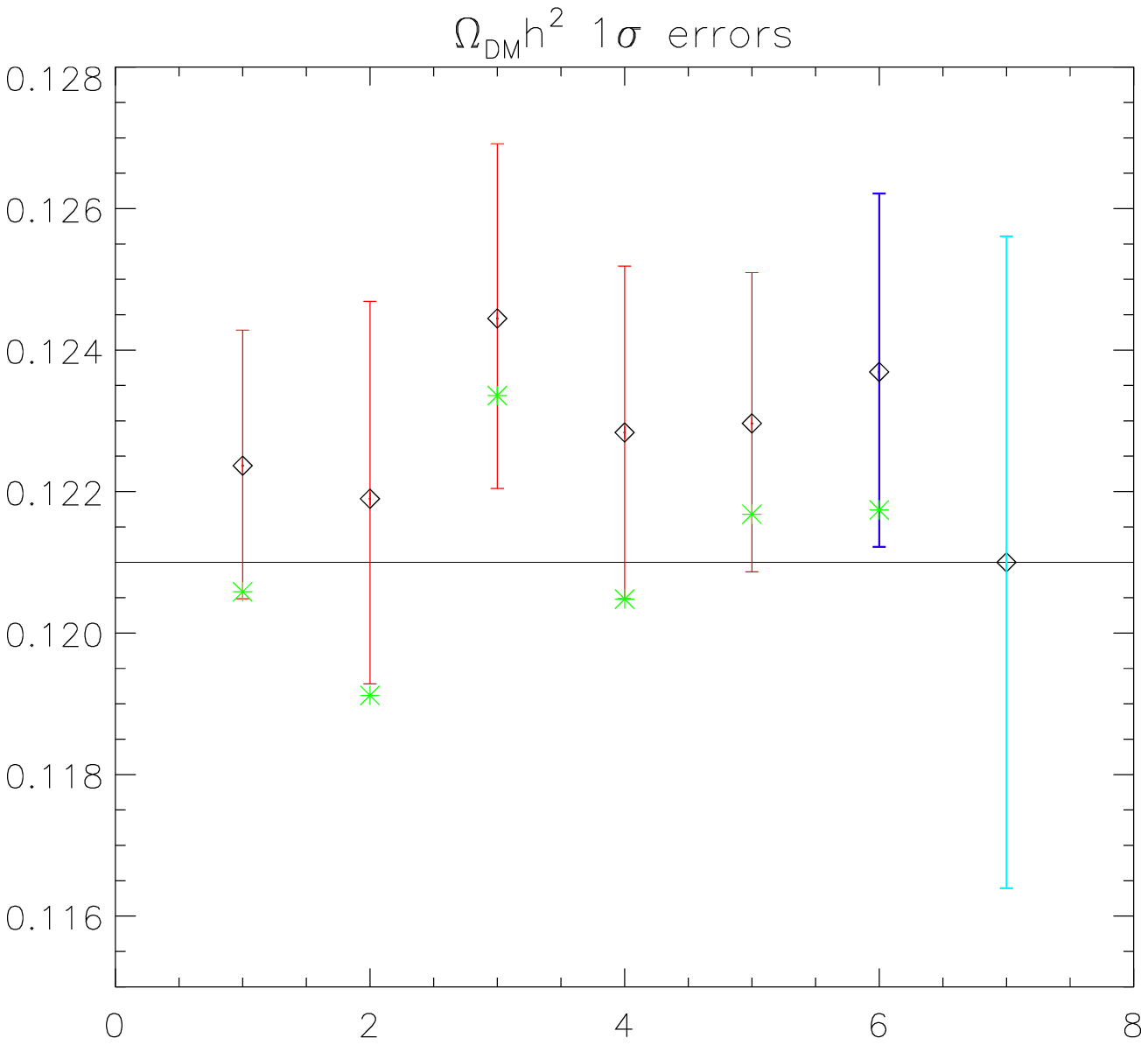}
\hfill
\caption{\label{fig2} Same as Figure \ref{fig2} but for
$\{f_{\nu}, \Omega_{dm} h^2\}$.
In addition to the mean values (diamonds), we show also the
best-fit values (green crosses) obtained from the Markov Chains.}
\end{center}
\end{figure*}

Figures \ref{fig1} and \ref{fig2} show the expected $1\sigma$
sensitivity of Planck to each of the eight parameters. So far, no lensing
information (i.e., $C_l^{dd}$ and $C_l^{Td}$) has been included in the analysis.
The errors in these Figures are obtained in three different
ways:
\begin{enumerate}
\item
First, we generate five independent realisations of $a_{lm}^P$
(see equation (\ref{realisation})) from the same cosmological
model, the parameter values of which (i.e., $\theta_i^0$)
are indicated by the horizontal lines.  The noise power spectrum corresponds to the expected Planck
sensitivity (using the three frequency channels with the lowest foreground
levels at 100, 143 and 217 GHz, see \cite{:2006uk} for details).
For each of our five Planck mock
data sets, we reconstruct the observed power spectra $\hat{C}_l^{PP'}$ as per
(\ref{reconstruction}), sample the likelihood (\ref{chieff}) with
CosmoMC, marginalise over all parameters but one, and plot the mean
values and $68\%$ confidence limits (1$\sigma$ errors).

\item
Second, we perform the parameter estimation not from a realisation of
the fiducial power spectra, but from the fiducial spectra
themselves. In other words, we directly sample the likelihood
(\ref{chieff}) with $\hat{C}_l^{PP'}$ set equal to the theoretical power
spectrum of the fiducial model plus the Planck noise spectrum, $\bar{C}_l^{PP'}$.
This amounts to considering an average over an infinite number of mock Planck
data sets. We sample again the likelihood with CosmoMC, marginalise
over all parameters but one, and plot the mean values and $68\%$
confidence limits.

\item
Third, we forecast the $68\%$ confidence limits using the Fisher matrix
formalism. For the computation of the
derivatives $\partial C_l^{PP'} / \partial \theta_i$, we choose a stepsize
of order $\Delta \theta_i \sim \sigma(\theta_i)$.
We centre the resulting error bars (which are symmetrical
by definition) on the fiducial values $\Theta^0$ in Figures \ref{fig1} and \ref{fig2}.
\end{enumerate}

\begin{figure*}[t]
\begin{center}
\hfill
\includegraphics[width=0.85\textwidth]{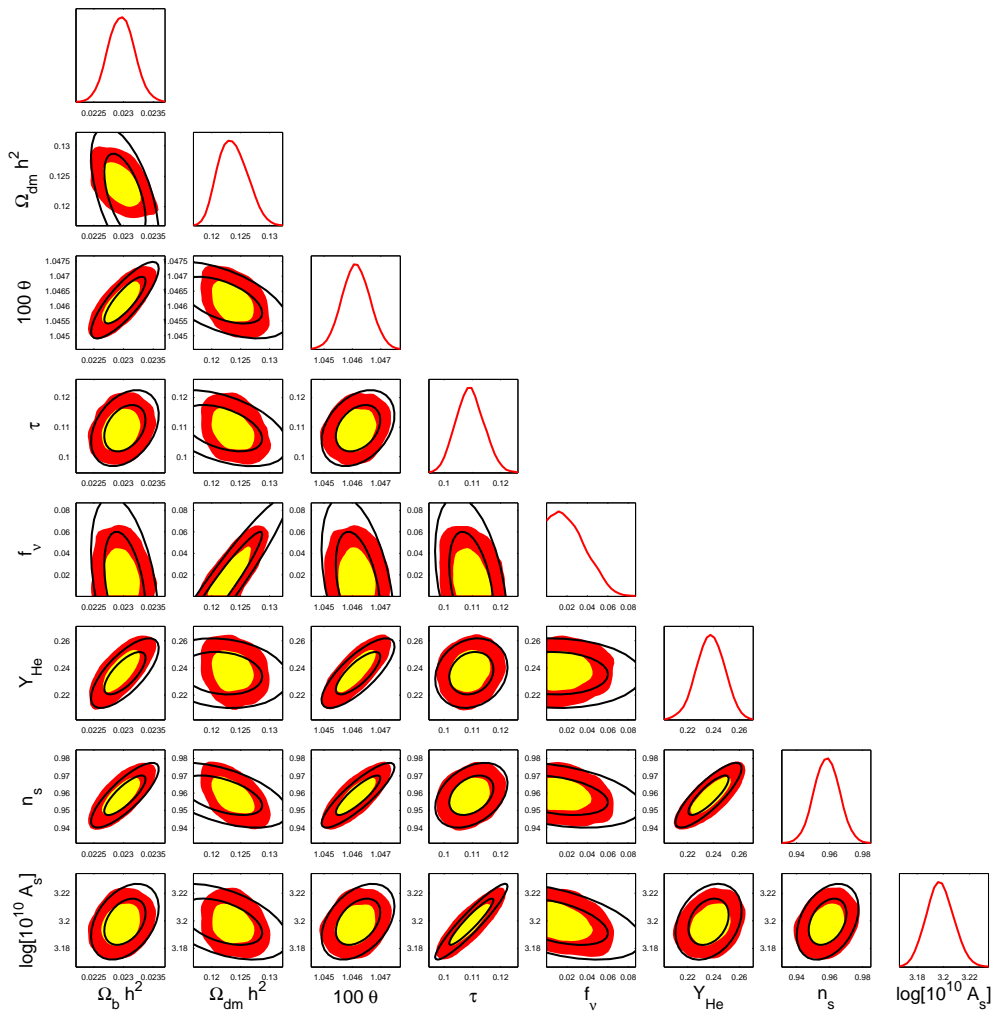}
\hfill
\caption{\label{fig3} Projected $68\%$ and $95\%$ confidence levels from
the Monte Carlo (colored/shaded) and the Fisher matrix (black lines)  methods,
for Planck without lensing extraction and the minimal, eight-parameter
$\Lambda$MDM model of section \ref{sec:minimal}.
The diagonal plots show the corresponding
marginalised probabilities for each cosmological parameter.
}
\end{center}
\end{figure*}

For all parameters but $\{ \Omega_{dm} h^2, f_{\nu}\}$, the three
methods are in good agreement and provide very similar errors
$\sigma(\theta_i)$. A comparison of the five independent mock data
results shows that the errors do not vary significantly from case
to case, and that the mean values are nicely distributed around
the fiducial value, with a typical dispersion in agreement with
the error bars.%
\footnote{Actually, one could object that the error
bars are a bit large with respect to the actual scattering of the
mean values, particularly for $\tau$. We attribute this to our
crude modelling of the galactic cut (see section \ref{sec:Mock}),
which leads to insufficient scattering of the mean values.
}
At this point, the Monte Carlo method based on the
fiducial spectrum (method (ii)) and the Fisher matrix approach (method (iii))
both appear to be
robust error forecasting techniques.

\begin{figure*}[t]
\begin{center}
\hfill
\includegraphics[width=0.85\textwidth]{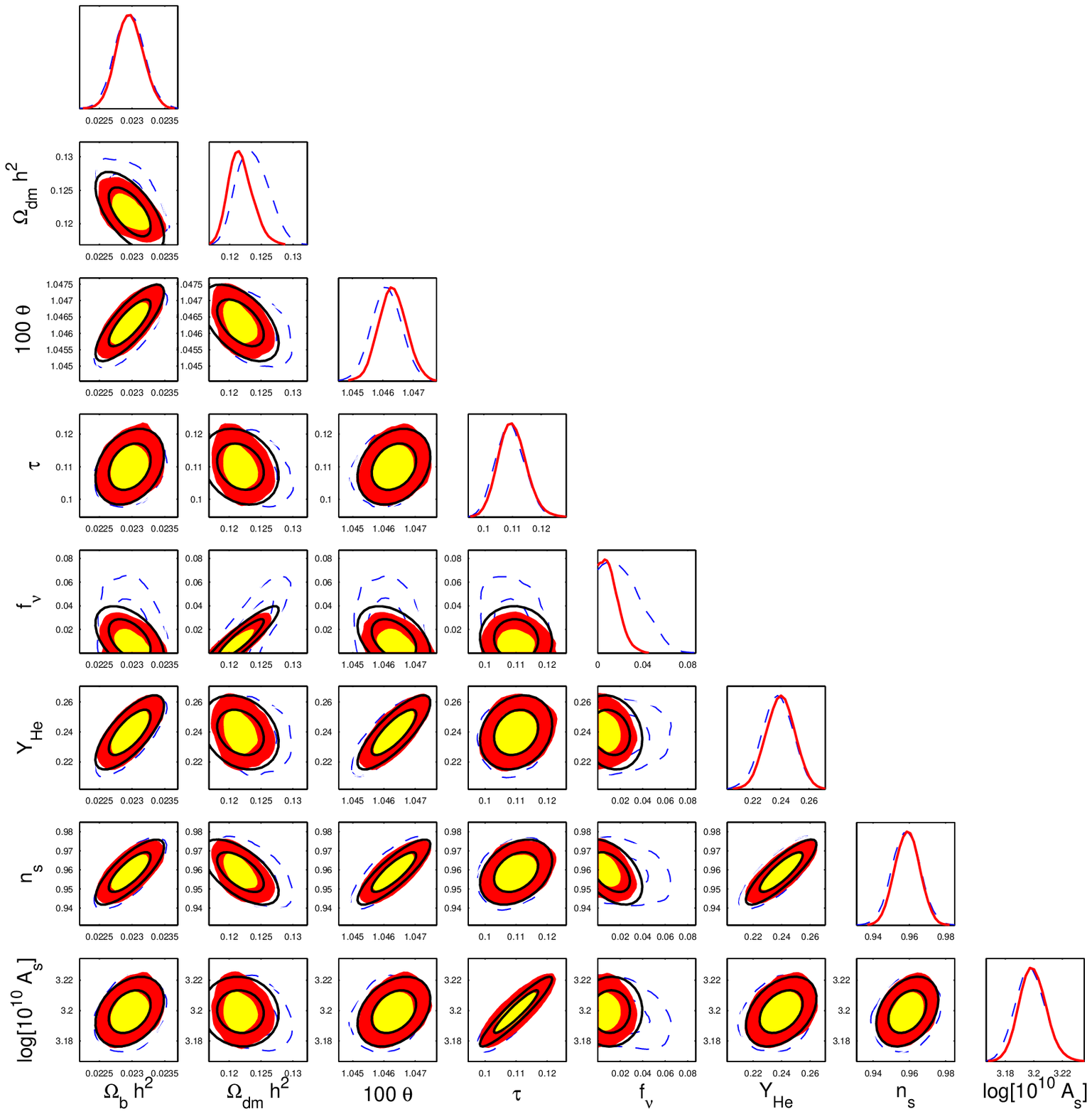}
\hfill
\caption{\label{fig4}
Projected $68\%$ and $95\%$ confidence levels from
the Monte Carlo (colored/shaded) and the Fisher matrix (black lines)
methods, for Planck with lensing extraction and the minimal, eight-parameter
$\Lambda$MDM model of section \ref{sec:minimal}.
The Monte Carlo results of the Figure \ref{fig3}
(for Planck without lensing extraction) are shown again for comparison
(blue dashed lines).  The diagonal plots show the corresponding
the marginalised probabilities for each cosmological parameter,
with (red) lensing extraction (red) and without (blue dashed).
}

\end{center}
\end{figure*}

The conclusions are drastically different for the parameters $\{
\Omega_{dm} h^2, f_{\nu}\}$. The mean values of the five mock data are
now distributed asymmetrically with respect to the fiducial values, and
the errors predicted by the Fisher matrix are bigger, roughly by
a factor two, than those obtained from CosmoMC.  Clearly, these
problems signal the existence of a non-Gaussian dependence of the
likelihood on $\{\Omega_{dm} h^2, f_{\nu}\}$.  A quick way to check
this is to plot the best-fit values%
\footnote{Note that, in general, the best-fit values are more
efficiently obtained from a
minimisation algorithm such as simulated annealing, than from a Monte
Carlo method.}
for each of our CosmoMC runs (green crosses in Figure \ref{fig2});
the best-fit values depart significantly from the mean values, as should be
the case for asymmetrical probability distributions. Moreover, the best-fit
values are scattered symmetrically around the fiducial values, confirming
the fact
that the maximum likelihood is an unbiased estimator of $\theta_i$.
(In contrast, the mean value becomes a biased estimator of $\theta_i$
as soon as the likelihood departs from Gaussianity with respect to
$\theta_i$.)

It is easy to understand why the likelihood is non-Gaussian with respect to
$f_{\nu}$: $f_{\nu}$ is confined to positive values only,
cutting the likelihood before it drops to zero. Future experiments or
combination of experiments will be confronted to this problem until
they have enough sensitivity for making a clear detection of a non-zero
neutrino mass.

The non-Gaussianity with respect to $\Omega_{dm} h^2$ can be
understood from an inspection of Figure \ref{fig3}, which shows
the two-dimensional likelihood contours for each pair of parameters
(CosmoMC is particularly convenient for obtaining such plots),
and the one-dimensional marginalised probabilities on the diagonal.
The $68\%$ (1$\sigma$) and $95\%$ (2$\sigma$) confidence contours
obtained from the Monte Carlo method (using the fiducial
spectra) are shown as the red/yellow
(dark/light) filled contours.  For all combinations not involving
$\Omega_{dm} h^2$ or $f_{\nu}$, remarkable agreement
with the black ellipses derived from the Fisher matrix can be seen.
(In practice, we obtain the Fisher matrix ellipses by inverting
the relevant $2\times2$ submatrix, which are then centred on
the best-fit
parameter values obtained from the Monte Carlo method).
From these plots, we see  that the parameter $f_{\nu}$ is correlated
mainly with $\Omega_{dm} h^2$.  This  correlation
means that any non-Gaussianity in
the $f_{\nu}$ probability will propagate to $\Omega_{dm} h^2$.
This is why the Fisher matrix
ellipses provide a poor approximation of the contours involving
$\Omega_{dm} h^2$ and/or $f_{\nu}$.

Better agreement with the Monte Carlo results could perhaps be achieved by
adjusting the stepsize when computating the derivatives
$\partial C_l^{PP'} / \partial f_{\nu}$
and $\partial C_l^P / \partial (\Omega_{dm} h^2)$ for the Fisher matrix.
However, as discussed in the section \ref{sec:Fisher},
there are no well-controlled methods for doing this unless
the full likelihood function is already available.

\begin{table*}[t]
\caption{\label{taball_1sig} Standard deviations (or 1$\sigma$ errors)
obtained from the Monte Carlo (MCMC) and the Fisher matrix methods, with
and without lensing extraction, for the minimal eight-parameter and
the extended eleven-parameter models. We also show the corresponding limits
for two important related parameters: the neutrino mass in units of ${\rm eV}$
 and the cosmological constant density fraction.}
{\footnotesize
\hspace{13mm}
\begin{tabular}{@{}c|cccc|ccccc}
\br
{\small method:}
& MCMC & Fisher & MCMC & Fisher & MCMC & Fisher & MCMC & Fisher\\
& \multicolumn{2}{c}{no lensing}
& \multicolumn{2}{c|}{lensing}
& \multicolumn{2}{c}{no lensing}
& \multicolumn{2}{c}{lensing}  \\
{\small param.} & & & & & & & & \\
\mr
$\Omega_bh^2$
&{\small 0.00022}&{\small 0.00023}&{\small 0.00020} &{\small 0.00021}
&{\small 0.00028}&{\small 0.00025}&{\small 0.00023} &{\small 0.00024}\\

$\Omega_{dm}h^2$
&{\small 0.0024} &{\small 0.0046} &{\small 0.0019} &{\small 0.0024}
&{\small 0.0092} &{\small 0.0073} &{\small 0.0046} &{\small 0.0048}
\\

$f_{\nu}$
& {\small 0.016} & {\small 0.034} & {\small 0.008} &{\small 0.011}
& {\small 0.030} & {\small 0.040} & {\small 0.009} &{\small 0.013}
\\

$\theta$
&{\small 0.0005}&{\small 0.0005}&{\small 0.0004}&{\small 0.0005}
&{\small 0.0013}&{\small 0.0011}&{\small 0.0010}&{\small 0.0010}
\\

$Y_{\rm He}$
& {\small 0.011} & {\small 0.011} & {\small 0.010}&{\small 0.010}
& {\small 0.020} & {\small 0.017} & {\small 0.017}&{\small 0.017}
\\

$\tau$
& {\small 0.0048}& {\small 0.0052}& {\small 0.0047}&{\small 0.0047}
& {\small 0.0053}& {\small 0.0062}& {\small 0.0050}&{\small 0.0049}
\\

$\log[A_s]$
& {\small 0.010} & {\small 0.011} & {\small 0.009}& {\small 0.009}
& {\small 0.013} & {\small 0.014} & {\small 0.012}& {\small 0.012}
\\

$n_s$
& {\small 0.007}& {\small 0.008}& {\small 0.007}&{\small 0.007}
& {\small 0.011} &{\small 0.011}& {\small 0.010}&{\small 0.010}
\\
\hline
w                & - & - & - & -
& {\small 0.49} &{\small 0.68}   &  {\small 0.23} & {\small 0.18}  \\
$N_{\rm eff}$        & - & - & - & -
& {\small 0.46} &{\small 0.27}   &  {\small 0.27} & {\small 0.26}    \\
$\alpha $        & - & - & - & -
& {\small 0.0090} &{\small 0.0087} & {\small 0.0075}& {\small 0.0077}
\\
\hline
$m_\nu$~(eV)
& {\small 0.19}  &{\small 0.45}   & {\small 0.09}  &  {\small 0.13}
& {\small 0.42}  &{\small 0.51}   & {\small 0.11}  &  {\small 0.15}   \\
$\Omega_{\Lambda}$
&{\small 0.02}   & {\small 0.05}  &  {\small 0.02} &{\small 0.02}
&{\small 0.12}   & {\small 0.17}  &  {\small 0.06} &{\small 0.06}  \\
\br
\end{tabular}
}
\end{table*}

The exercise is repeated in Figure \ref{fig4}, but now with the
inclusion of lensing information.  Here, we do not generate independent
realisations of the data,
since the Monte Carlo method utilising the fiducial power
spectrum performs equally well for the purpose of error forecasting,
as demonstrated in section \ref{sec:minimal}.
Lensing extraction is particularly useful for constraining physical
quantities that affect the late evolution of cosmological
perturbations~\cite{Hu:2001fb,Kaplinghat:2003bh,Lesgourgues:2005yv}.
These quantities include dark energy (or a cosmological constant) which
reduces the growth of matter perturbations at low redshifts, and
neutrino masses, which also suppress this growth on small scales.

Since the present parameter basis does not include $\Omega_{\Lambda}$,
the effect of dark energy may be difficult to discern in Figure \ref{fig4}.
However, the significant sharpening of the $f_{\nu}$ probability
distribution is clearly visible.  Lensing probes the matter perturbations
in a more direct way (than does the CMB alone).  This allows for
a better determination of $f_{\nu}$ through the neutrino's free-streaming effect
on the matter power spectrum.  The degeneracy with $\Omega_{dm} h^2$ is also reduced
as a consequence.  Thus, the likelihood function is now much closer to a
multivariate Gaussian, and the Fisher matrix appears to provide satisfactory results,
as can be seen from a comparison of the Fisher matrix ellipses with the
actual, Monte Carlo contours in Figure \ref{fig4}.

In Table \ref{taball_1sig}, we provide the numerical values of the
1$\sigma$ errors obtained from the Monte Carlo and the Fisher matrix
methods, with and without lensing extraction.  We show also the
corresponding limits for two related parameters, the neutrino mass and
the cosmological constant density fraction.  Just as for $\Omega_{dm}$
and $f_{\nu}$, the 1$\sigma$ errors for these two quantities are very
discrepant between the two cases, since in absence of lensing extraction
the Fisher matrix overestimates the $m_{\nu}$ error by a factor 2.4,
and that on $\Omega_{\Lambda}$ by 2.5.

\begin{figure*}[t]
\begin{center}
\hfill
\includegraphics[width=0.99\textwidth]{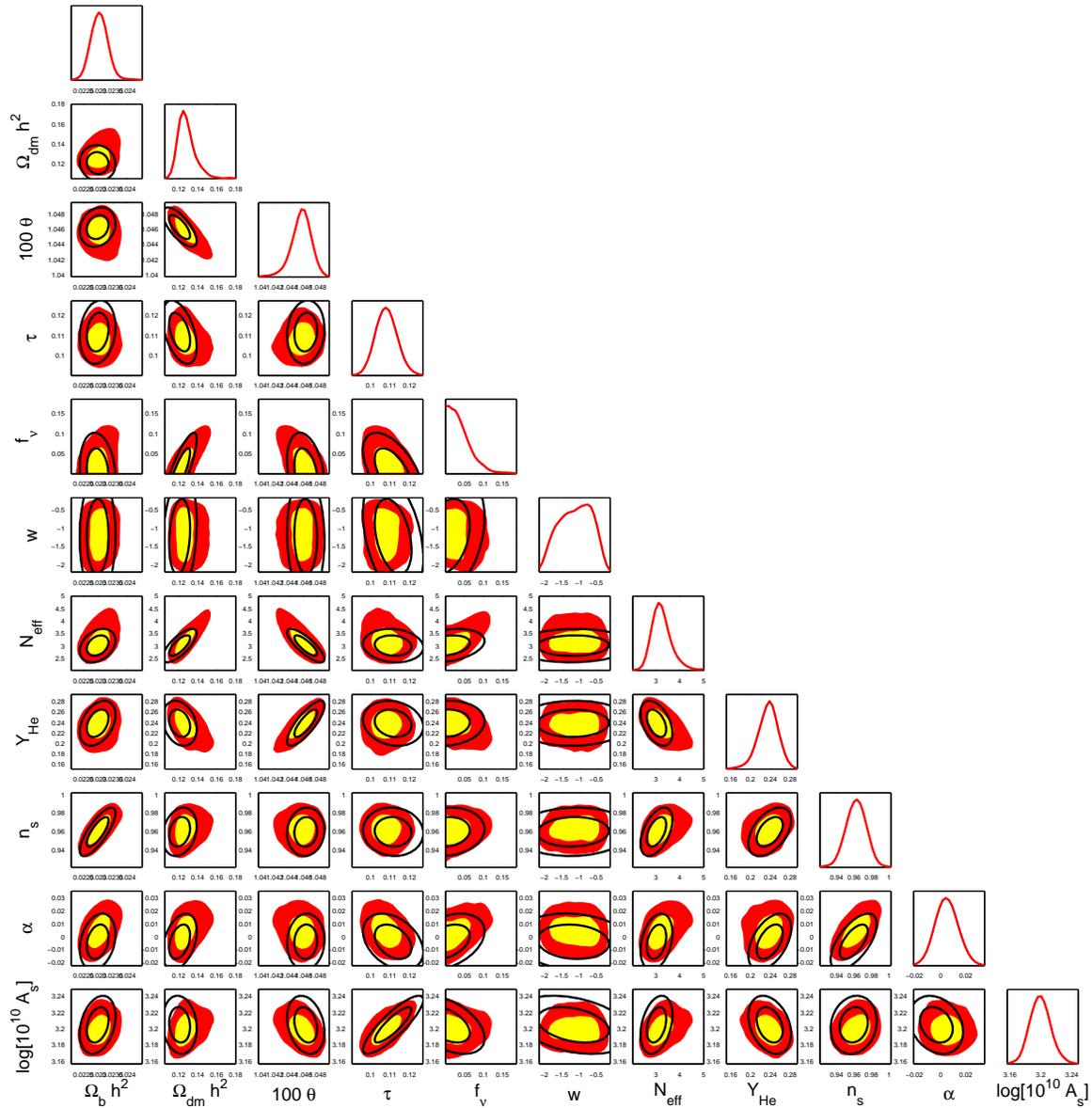}
\hfill
\caption{\label{fig5}
Projected $68\%$ and $95\%$ confidence levels from
the Monte Carlo (colored/shaded) and the Fisher matrix (black lines)  methods,
for Planck without lensing extraction and the extended, eleven-parameter
$\Lambda$MDM model of section \ref{sec:non-minimal}.
The diagonal plots show the corresponding
marginalised probabilities for each cosmological parameter.}
\end{center}
\end{figure*}

\section{Results including non-minimal parameters}
\label{sec:non-minimal}

We now study a non-minimal cosmological model with three extra
parameters: a (constant) dark energy equation of state $w$, a
running spectral index $\alpha$, and the effective number of
massless neutrinos $N_{\rm eff}$ (or the number of
relativistic fermionic degrees of freedom). Explicitly,
our model consists of one massive neutrino responsible for the
hot fraction of dark matter $f_{\nu}$,
plus a relativistic density attributed to $(N_{\rm eff}-2)$ massless neutrinos.
Thus, the total number of independent parameters is now eleven.

\begin{figure*}[t]
\begin{center}
\hfill
\includegraphics[width=0.99\textwidth]{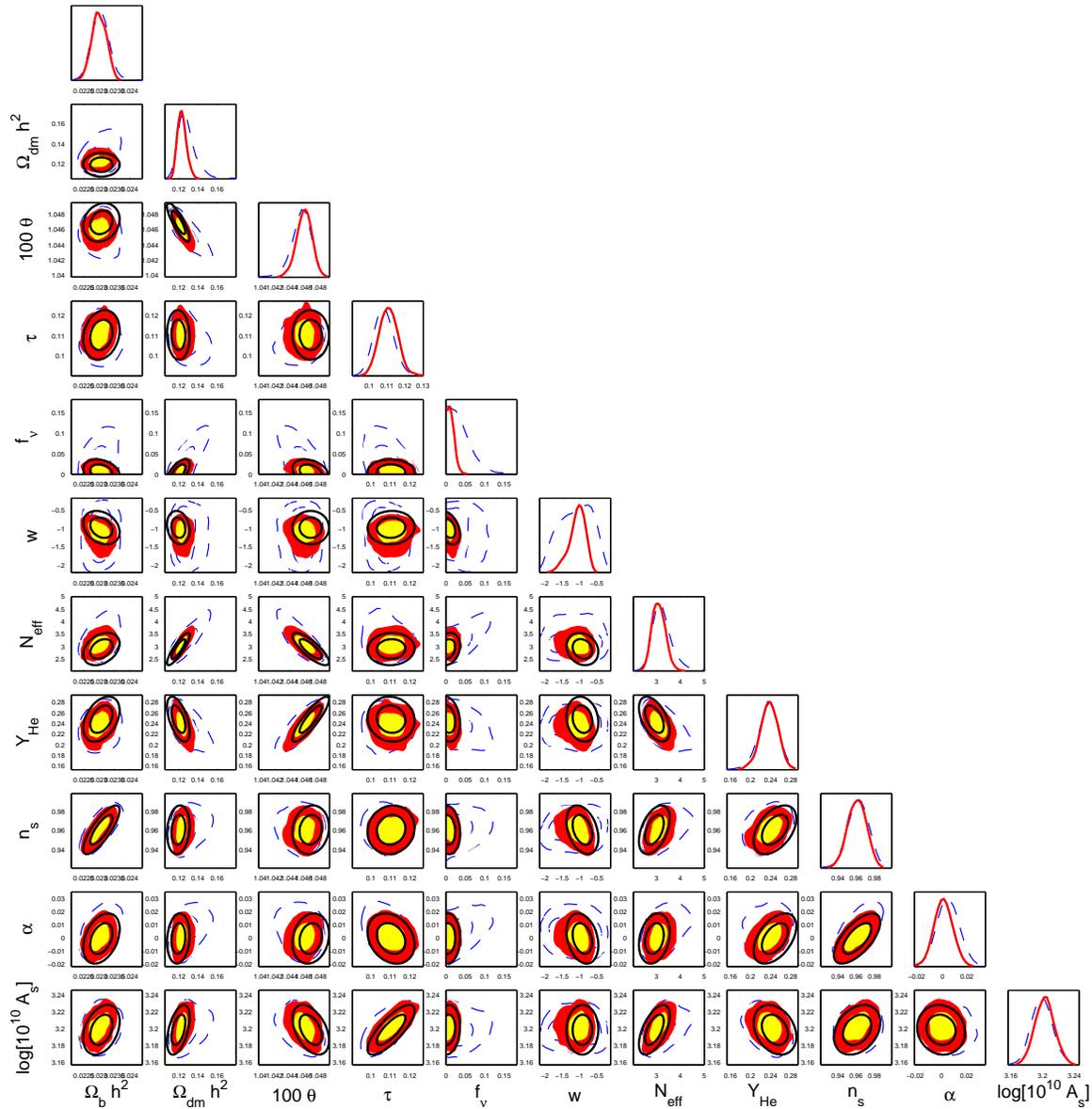}
\hfill
\caption{\label{fig6}
Projected $68\%$ and $95\%$ confidence levels from
the Monte Carlo (colored/shaded) and the Fisher matrix (black lines)
methods, for Planck with lensing extraction and the extended, eleven-parameter
$\Lambda$MDM model of section \ref{sec:non-minimal}.
The Monte Carlo results of the Figure \ref{fig5}
(for Planck without lensing extraction) are shown again for comparison
(blue dashed lines).  The diagonal plots show the corresponding
the marginalised probabilities for each cosmological parameter,
with (red) lensing extraction (red) and without (blue dashed).}
\end{center}
\end{figure*}

Without lensing extraction, we saw in section \ref{sec:minimal}
that the minimal eight-parameter model was already poorly
constrained as a consequence of a mild $\Omega_{dm} h^2,f_{\nu}$
degeneracy. In the present eleven-parameter model, the situation
worsens, mainly because of another parameter degeneracy between
$N_{\rm eff}$, $\Omega_{dm} h^2$, $\theta$ and $f_{\nu}$ (see e.g.,
\cite{Hannestad:2003xv,HR04,Crotty:2004gm,Dodelson:2005tp}). These
degeneracies manifest themselves in Figure \ref{fig5} as very
elongated contours, leading clearly to a non-Gaussian likelihood
with respect to many cosmological parameters. So, it is not
surprising to find that the Fisher matrix is a poor approximation
in many cases.

As expected, the inclusion of lensing extraction offers vast
improvements in the determination of $w$ and $f_{\nu}$.  Consequently,
the correlations between $\Omega_{dm} h^2$ and $f_{\nu}$ and
between $N_{\rm eff}$ and $f_{\nu}$ essentially disappear (see Figure
\ref{fig6}).  A reduction of the degeneracy between $\Omega_{dm}
h^2$, $\theta$  and $N_{\rm eff}$ can also be seen, although some correlation
between the two parameters remains (since it is possible to vary
these parameters simultaneously without changing the epoch of
matter--radiation equality).  In general, the ``lensing'' contours
in Figure \ref{fig6} are much more elliptic than their ``no lensing''
counterparts, indicating that the likelihood is better fitted
by a multivariate gaussian.

Table \ref{taball_1sig} shows the numerical values of the 1$\sigma$
errors for the eleven-parameter model, with and without lensing
extraction, obtained from the two forecast methods.  In the case
without lensing extraction, the Fisher matrix still overestimates the
error on $f_{\nu}$ and $\Omega_{dm} h^2$, as well as that on $w$. For
$N_{\rm eff}$, the likelihood is strongly non-gaussian (with large
skewness) and the Fisher matrix underestimates the 1$\sigma$
error by a factor 1.7.
The discrepancies are even stronger when one looks at the 2$\sigma$
errors.

\section{Discussion} \label{sec:Conclusions}

We have studied error forecasts on cosmological parameters from
the Planck satellite using two different methods. The first is the
conventional Fisher matrix analysis in which the second derivative
of the parameter likelihood function at the best fit point is used to
calculate formal 1$\sigma$ errors on the parameters, as well as the
parameter correlation matrix.  The second is to use Markov Chain Monte
Carlo methods such as CosmoMC on synthetic data sets.  This is the
same method normally used to extract parameters from present data.

The Monte Carlo method has many advantages over the Fisher matrix
approach.  While the Fisher matrix uses only information at the
best fit point and assumes the likelihood function to be Gaussian
with respect to the model parameters,  the use of Monte Carlo
methods in conjunction with synthetic data maps out the true
likelihood function for the given model realisation.

In this paper, we have shown that the likelihood function can be
highly non-Gaussian, particularly with respect to the neutrino mass
and the dark matter density, and, as a result, the CosmoMC analysis
can give results that are very different from its Fisher matrix
counterpart.  For prospective Planck data without lensing
extraction and assuming a simple eight-parameter model,
the difference in the projected errors
can be as large as a factor of two or more for the said parameters.
This indicates that in such cases the Fisher matrix method does not
provide a reliable estimation. Including additional data such as
CMB lensing information breaks some of the parameter degeneracies,
and makes the likelihood more Gaussian.  The two methods thus become
more compatible.

On the other hand, adding more cosmological parameters (all of
which are physically motivated) leads to new parameter
degeneracies, and generally worsens the agreement between the two
forecast methods.  For the eleven-parameter model studied here, we
find a difference for the neutrino mass fraction of 45\% between
the two methods at the 68\% level, even with the inclusion of CMB
lensing. The conclusion is that for some parameters, even with the
very high precision of future data the likelihood function will
not be sufficiently Gaussian to yield a reliable estimate of the
precision with which the parameter can be measured using the
Fisher matrix approach.

Given that Monte Carlo analysis of simulated data sets is
computationally feasible with present computers, we propose that
future error forecast analyses should employ this method rather
than the Fisher matrix analysis. This will have the added
advantage that the same parameter extraction pipeline can be used
on real data as it becomes available.
The present work only includes CMB data simulated to mimic Planck.
However, the method can be easily generalised to include other
data sets such as future weak lensing and baryon acoustic
oscillation surveys.

\section*{Acknowledgements}
We would like to thank Sergio Pastor and Massimiliano Lattanzi
for useful discussions on this work.

\section*{References}


\end{document}